\documentclass{article}

\newcommand{\be}{\begin{equation}}
\newcommand{\ee}{\end{equation}}
\newcommand{\bea}{\begin{eqnarray}}
\newcommand{\eea}{\end{eqnarray}}
\newcommand{\nn}{\nonumber}
\newcommand{\lb}{\left[}
\newcommand{\rb}{\right]}
\newcommand{\ac}{\mathcal{A}}
\newcommand{\hac}{\hat\mathcal{A}}

\begin{document}
\begin{titlepage}

\title{Note on the structure of constraint algebras}

\author{M. Stoilov\\
{\small\it Institute of Nuclear Research and Nuclear Energy,}\\
{\small\it Blvd. Tzarigradsko Chausse\'e 72, Sofia 1784, Bulgaria}\\
{\small e-mail: mstoilov@inrne.bas.bg}}

\maketitle

\begin{abstract}
It is shown that when the gauge algebra is with root system the 
canonical Hamiltonian  commutes with the constraints.
Two other simple propositions concerning gauge fixing
are proved too.
\end{abstract}

\end{titlepage}

There is a vast literature on systems with constraints.
Here we follow the Hamiltonian approach in which any constraint system 
is characterized by its Hamiltonian $H$ and constraints
$ \varphi_a =0,\;\;a=1,\dots,m$ (all of them functions in the phase space).
A thorough treatment of the subject could be found in  \cite{Hen}.
In the present paper we shall consider the simplest and common case  of
order one models with first class  (Bose) constraints.
These requirements  mean that 
the following Poisson bracket relations hold:
\bea
 \lb\varphi_a, \varphi_b\rb & = &C_{abc}\varphi_c, \label{algi}\\
\lb H, \varphi_a\rb & = & U_{ab}\varphi_b,.\label{algii}
\eea
with coefficients $U_{ab}$ and $C_{abc}$ 
independent of dynamical variables
(order one requirement).
The coefficients $C_{abc}$ are the structure constant of the 
algebra of gauge symmetry $\ac$ in the model.
As we see from eqs.(\ref{algi}) the generators of this algebra are the first class
constraints $\varphi$.

A note should be added at this point.
In the Hamiltonian approach the Hamiltonian is not uniquely determined.
It is always possible to add to it a combination of constraints with arbitrary coefficients 
(these are the so called weakly zero terms).
The origin of this ambiguity is the fact that in
 any calculation only the  total Hamiltonian $H_t$
\be H_t=H+\lambda_a\varphi_a,\label{htot}\ee
appears.
Here  $\lambda_a$ are new (arbitrary) variables --- the Lagrange multipliers.
Formula (\ref{htot}) shows that any weakly zero term in the Hamiltonian
can be absorbed in the 
$\lambda_a\varphi_a$ term after redefinition 
(with unity Jacobian) of the Lagrange multipliers.
So, it is convenient to introduce the  so called canonical Hamiltonian  $H_c$
which  is  Hamiltonian of the system with all weakly zero terms removed:
\be H_c=H \vert_{\varphi =0}\label{hcan}\ee
As usual we suppress the subscript `c' in $H_c$ and throughout 
the whole paper (including eqs.(\ref{algii})) $H$ denotes the canonical Hamiltonian.

Our aim here is to prove three propositions:

{\bf Proposition 1} If the  gauge algebra $\ac$ defined by eqs. (\ref{algi})
 is a Lie algebra with root system, then
the coefficients $U_{ab}$ in eq.(\ref{algii}) are  zeros.

{\bf Proof}\\
Eqs.(\ref{algii}) originate from  the requirement that the time
evolution preserves the gauge algebra $\ac$ generated by constraints $\varphi$. 
But it also asserts that $\varphi$ and $H$  form closed algebra 
which we shall denote by $\hac$.

There are two possibilities for the algebra $\hac$: 
\begin{itemize}
\item it can coincide with the gauge algebra $\ac$ (or with its universal envelope), or
\item it can be with one generator larger.
\end{itemize}
In the first case $\hac$  is trivially closed.
However, in this case
$H$ has to be a combination of the generators $\varphi_a$,
and so is in fact zero  because all weakly zero terms are removed from 
the canonical Hamiltonian $H$.
Therefore, in this case $U_{ab} = 0$.

Consider now  the case when $\hac$ is
larger than the gauge algebra.
Here we shall use the requirement that $\ac$ is a Lie algebra with root system.
For such algebra
there are three possibilities for $H$. It can be
\begin{itemize}
\item  a new independent  step operator, or
\item a new independent Cartan operator,  or
\item an independent central element of $\ac$.
\end{itemize}
However,  $\hac$ cannot be closed if $H$ is a new step operator --- 
there has to be at least one more step operator (the conjugated one) 
and also the corresponding Cartan element, both of which cannot be  
part of the gauge algebra (and thus of  $\hac$ also). 
Same arguments show that $\hac$ is not closed 
if $H$ is an independent Cartan operator. 
In this case  we have at least one new root, so we
have to add also at least one couple of new, independent step operators.
The only possibility we are left with is that $H$ is in the center of $\ac$.
But this means that $U_{ab} = 0$. {\bf QED}

A proper treatment of  any gauge model requires supplementary gauge 
fixing conditions. 
The only requirement on these conditions is their gauge noninvariance for
any gauge transformation.
Our next propositions concern two possible gauge choices ---
first, when gauge conditions are functions of the phase space variables and 
second, when they depend on Lagrange multipliers.

{\bf Proposition 2} If  the gauge conditions depend on phase space variables only
it is possible to add weakly zero terms to the canonical Hamiltonian, such that
the new Hamiltonian has zero Poisson bracket with the gauge conditions.

{\bf Proof}\\
The gauge transformation of any function in phase space is 
generated by constraints through the Poisson bracket relations.
Therefore, functions $\chi_a$  of canonical coordinates and their momenta
can be used as gauge conditions provided the operator $\Delta$ 
\be
\Delta_{ab} \equiv \lb\chi_a,\varphi_b\rb \label{del}
\ee
is invertible and hereafter we shall suppose that this is fulfilled.

Let us introduce  notations $y_a\equiv \lb H, \chi_a\rb $
and $D_{ab} : \;\;D_{ab}f_b=\varphi_b\lb \chi_a, f_b\rb $.
Our task is for $y \ne 0$ to find a vector $\alpha$, such that 
\be
\lb H + \alpha_a \varphi_a, \chi_b\rb=0 \label{los}.
\ee
A particular solution of this system of first order differential equations
can be found as a formal series:
\bea
\alpha            &=& \sum_{n=0} \alpha^{(n)}\nn\\
\alpha^{(0)} &=& \Delta^{-1}  y\nn\\
\alpha^{(n)} &=& -\Delta^{-1}\lb\chi,\alpha^{(n-1)}\rb\varphi.\label{mno}
\eea
The solution given by eqs.(\ref{mno}) represents the series decomposition
of $(\Delta + D)^{-1}$ in the neighborhood of the "point" $\Delta$.
{\bf QED}
 
{\bf Proposition 3} The most general form of gauge condition, involving the 
Lagrangian multipliers only is
$
\lambda_a=\lambda^0_a, 
$
where $\lambda^0_a$ are some constants.

{\bf Proof}\\
Suppose that we have a vector function $f$ of $\lambda$ such that
$f(\lambda)=0$ fixes gauge freedom completely, i.e.
\be
\delta_\epsilon f \ne 0 \;\;\;\forall \epsilon \ne 0. \label{varf}
\ee
Here $\delta_\epsilon$ is the operator of infinitesimal gauge
transformation with parameter $\epsilon$. 
However, any variation of $f$ is given by:
\be
\delta f = \frac{\partial f}{\partial \lambda} \delta\lambda.
\ee
Provided that 
$
\lambda_a = {\rm constants}, 
$
is a good gauge condition (which is usually supposed),
then the operator of gauge transformations acting on 
the Lagrange multipliers  has no zero modes and so,
eq.(\ref{varf}) is fulfilled only provided 
the matrix $\frac{\partial f}{\partial \lambda}$ also has no zero modes.
Therefore, 
 dim$(f)=m$ and
\be
{\rm det}\vert\frac{\partial f}{\partial \lambda}\vert \ne 0
\ee
in the vicinity of the point $\lambda^0$ such that $f(\lambda^0)=0$.
As a result $f(\lambda)$ are invertable functions.
Their inverse we shall denote by $\bar\lambda(f)$.
The expression 
\be
\lambda=\bar\lambda(0)=\lambda^0 
\ee
gives the equivalent transcription of
the gauge condition $f(\lambda)=0$. {\bf QED}

\end {document}